A STUDY ON THE EVOLUTION OF A COMMUNITY POPULATION BY CUMULATIVE AND FRACTIONAL CALCULUS APPROACHES

F.Büyükkılıç*, Z.Ok Bayrakdar , D.Demirhan
Department of Physics, Faculty of Science, EgeUniversity, Izmir, 35100 Bornova, Turkey

**Abstract**

Nowadays, in our globalized world, the local and intercountry movements of population have been increased. This situation makes it important for host countries to do right predictions for the future population of their native people as well as immigrant people. The knowledge of the attained number of accumulated population is necessary for future planning, concerning to education, health, job, housing, safety requirements, etc. In this work, for updating historically well known formulas of population dynamics of a community are revisited in the framework of compound growth and fractional calculus to get more realistic relations.

Within this context, for a time t, the population evolution of a society which owns two different components is calculated. Concomitant relations have been developed to provide a comparison between the native population and the immigrant population that come into existence where at each time interval a colonial population is joined. Eventually at time t, the case where the native population becomes equal to immigrant population is discussed. Moreover, the differential equation of a probability function for the equilibrium state of accumulative population is obtained. It is seen that the equilibrium solution can be written as a series in terms of Bernoulli numbers. In the calculation of population dynamics, Mittag-Leffler(M-L) function replaces the exponential function.

**1. Introduction**

The economical and regional differences within a country as well as intercountries, create an irresistable force field for the mobility of human populations. With the developments in the communication technology, industry, transportation, tourism and with the acceleration of the transmission of sound and image around the world, eventually, the world has become well known for her inhabitants. Furthermore, by virtue of regional and local disagreements, wars, drought, famine, unemployment and diseases, human communities are forced to migrate in colonies.

Interaction of the human communities with their environments are of vital importance for biodynamics. Decrease/growth of human population has a strategic importance for the future of the countries. Survival of the states is possible with a stable population evolution, in another words by population balance. In order to cope with the problems of the future concerning to the education, health, sheltering, transportation, etc., future population needs to be known. In this circumstances, it becomes important to follow the growth and diminution of populations as well as the compatibility of population with environment. In this regard, population growth rate is a deterministic parameter for societies.

Simple exponential , unlimited growth model of population which is introduced by T.Malthus [1] is inadequate to describe a realistic population growth processes. The more realistic model of population growth, which contains restriction of the carrying capacity of environmentis introduced by P.F.Verhulst[2]. The equation of this model is known as logistic equation which is a special form of a nonlinear Bernoulli equation.

Time evolution of population dynamics can be handled as discrete or continuous processes[3-6]. It is obvious that population dynamics interacts with its environment where it evolves. Therefore, population dynamics as a complex process evolve in a fractal medium and within a Markovian manner. In this perspective, cumulative growth and diminution and fractional calculus approaches are suitable formulations to describe the complex dynamics of population[7-9]. In our cumulative and fractional calculus approaches, present and past or present and future states of the systems are interrelated that involve memory effects . In view of historical perspective, in a broad sense it can be named as "philosophy of Fibonacci". This universal of noted philosophy and its instruments, namely fractional calculus and cummulative approaches could be adopted to different fields of social, economical, statistical sciences. In this framework, Matthew effect, preferential attachment, Zipf's law, self organisation, human cooperation, dynamics of evolution, etc[10-13] could be justified.

In this work, population dynamics is handled as a cumulative growth/diminution process where time is considered as a discrete variable. This study is oulined as follows. In section1, population evolution is handled as a restricted and cumulatively growing process where the environment-system interaction is taken into account with an operator parameter $\alpha$. In section2, discrete logistic equation is obtained by departing from the cumulative growth where time is taken as a discrete parameter, with the restriction concerning to growth rate $a$ due to carrying capacity. For the continuous time, the logistic equation is rewritten where Bernoulli type nonlinear equation of population growth is obtained. In section3, considering that a colonial population $P$ joins to the native population in each step, the accumulative population is calculated. The cases for the accumulative population to be negative, zero and positive are investigated in terms of the quantities of native and colonial population components. In section4, by passing from discrete time variable to continuous one, conditions of equilibrium state are considered where primary population and colonial population are equal. The solution of the equation for the equilibrium state of colonial population is expressed in terms of Bernoulli numbers. In section5, it is asserted that the solution of the rate equation within fractional calculus and cumulative approaches are equivalent. In section 6, two component population dynamics is investigated by fractional calculus that involves the order of the fractional derivative $\alpha$ which takes into account the environmental effect. In section7, a hybrid solution of logistic equation of population is obtained within the context of cumulative approach. The solutions are illustrated by the concomitant figures.

## 2. Cumulative Growth of a Population

Let us formulate the growth process of a community population that has one component. The population at time t is $N(t)$ then after a time interval $\Delta t$ the population $N(t + \Delta t)$ of a discrete growth process in Euclidean space is defined as:

$$N(t + \Delta t) = (1 + a\Delta t)N(t) \tag{1}$$

where $a$ is growth rate which is independent of time[14,15]. If the environmental influences are considered, in which the dimension of the space is fractal and memory considerations are non-Markovian, then the evolution operator can be taken as

$$\phi = (1 + a\Delta t)^\alpha \tag{2}$$

where, the parameter $\alpha$ is in interval $0 < \alpha \leq 1$. In the following sections, parameter $\alpha$ is related to the order of the fractional derivative within the framework of fractional calculus.

The population at the $n$th stage is

$$N_n = \phi N_{n-1} = \phi^n N_0 = [(1 + a\Delta t)^\alpha]^n N_0 \tag{3}$$

where $N_0$ is initial population[14,15]. The continuous solution of the algorithm (3) in view of the cumulative growth approach is deduced in terms of M-L function as

$$N(t) = N_0 E_\alpha(at^\alpha) \tag{4}$$

[14-17], here, for $z = at^\alpha$, M-L function is defined as

$$E_\alpha(z) = \sum_{k=0}^{\infty} \frac{z^k}{\Gamma(\alpha k + 1)}. \tag{5}$$

Time dependence of the population growth is given in Fig.1. For a cumulative growing process, solution; Eq.(4) is deduced depending on a parameter $\alpha$. For different values of $\alpha$, Fig.1 is plotted. It is observed that for the case $\alpha = 1$ it is reduced to standard exponential growth function.

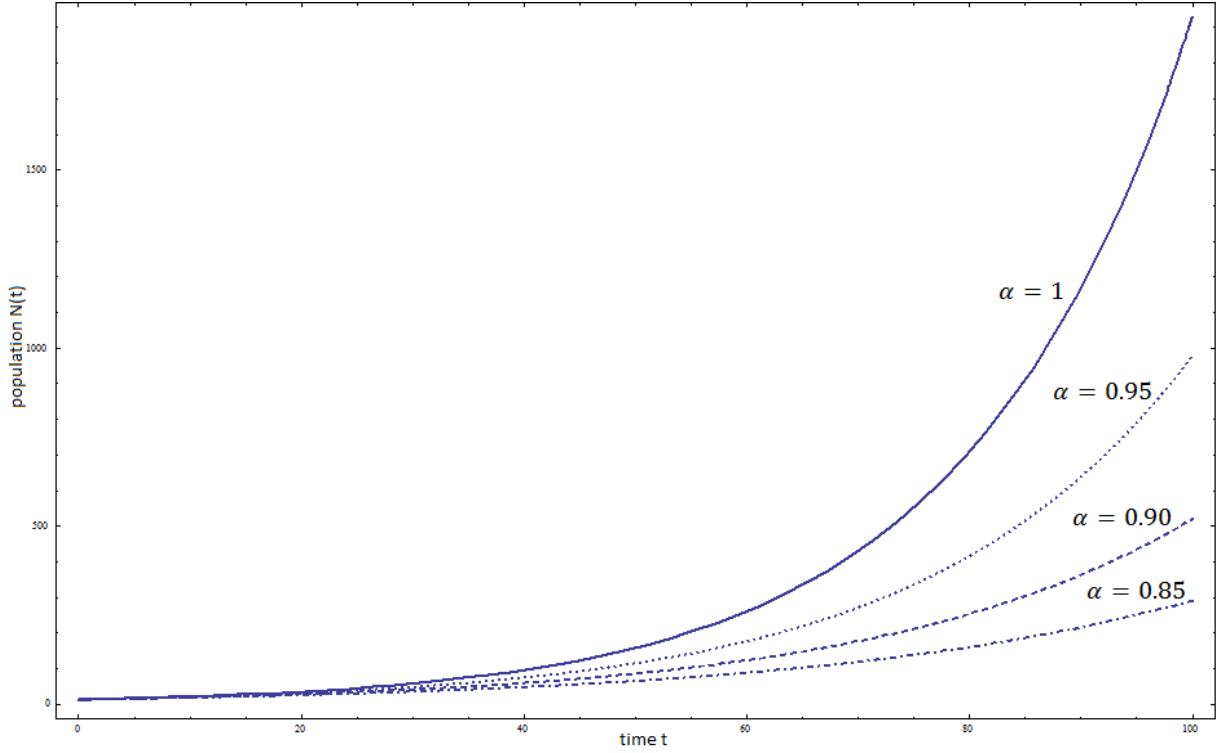

Fig.1.The plot of Eq.(4) for $N_0 = 13$ units, growth rate a=0.05, where $\alpha = 1$ (bold), $\alpha = 0.95$ (dotted), $\alpha = 0.9$ (dashed), $\alpha = 0.85$ (dotdashed) are presented

From Fig.1, it can be observed that the growth of population as a cumulative evolution is analysed for growth rate a=0.05 and different values of $\alpha$. It is concluded from Fig.1 that the environment has a retarding and slowing effect on the cumulatively growing population number which is imposed by parameter $\alpha$ with respect to standard exponential growth.

For Eq. (4), M-L function has the following limit values:

$$\lim_{t \to 0} E_\alpha (at^\alpha) = 1 + \frac{at^\alpha}{\Gamma(1 + \alpha)} + \cdots \approx exp\left[\frac{a}{\Gamma(1 + \alpha)} t^\alpha\right] \quad (6a)$$

$$\lim_{t \to \infty} E_\alpha (at^\alpha) \approx \frac{1}{a\Gamma(1 - \alpha)t^\alpha} . \quad (6b)$$

From these equations, it can be seen that solution (4) is an extrapolation between the exponential and the inverse power law, depend on the initial value $N_0$[9, 16-19].

As an alternative, a simple growth formula which contains the environmental effects due to $\alpha$, it can also be written

$$N_n = \phi^n N_0 = [(1 + a\frac{t}{n})^n]^\alpha N_0 . \quad (7a)$$

For sufficiently large $n$

$$N(t) = \lim_{n\to\infty}[(1+a\frac{t}{n})^n]^\alpha N_0 = N_0\exp(\alpha at) \qquad (7b)$$

can be written. Introducing $z = at$, for population at time $t$, the following relation

$$N(z) = N_0\exp(\alpha z) \qquad (8)$$

is obtained.

As a special case, for the values $\alpha z = \alpha at \ll 1$, this solution can be expanded into a series then

$$N(z) \approx \left(1 + \alpha z + \frac{(\alpha z)^2}{2!} + \cdots\right)N_0 \qquad (9)$$

can be written. The relation given by (9) is a very simple expression that contains the paremeter $\alpha$ which is encountered in fractional calculus. Despite long time $t$(i.e. for large $z$), Eq.(9) converges, where $\alpha$ and $a$ remain small.

Here, the point we must emphasise is that if the steps are increased, the function that exhibit growth became independent of the number of steps, namely, the process becomes a continuous process instead of a discrete in terms of time.

But, an unlimited growth model is not a realistic and a sufficient scenario for the description of population growth. Therefore with regard to carrying capacity of the environment, the population growth should be restricted.

## 3. Revision of the restriction of the population growth model

In section1, without any limitation except parameter $\alpha$, the concomitant relations which depend on time and growth rate was obtained for a community population. When these relations for population are compared with the real population growths, it becomes necessary to introduce some restrictions on population growth rate. In population dynamics which is the subject of this work, the main restrictive parameters are the sources which are necessary for maintanance of life. The biodynamical system in the course of time leads to an equilibrium state and it is provided by imposing constraints on the growth rate. In this section, the logistic equation and logistic map that contains limitations for population are handled within the framework of cumulative growth.

In previous works of the present authors[14,15], the population at$(n + 1)$th stage is given by a cumulatively evolving process for $\alpha = 1$and is described by

$$N_{n+1} = (1 + a\Delta t)N_n. \qquad (10)$$

This equation obviously can not describe realistically a population growth. In another words, it can not increase continuously forever. The decisive effect of the environment such as food deficiency in proportion to population, war, disease, drough, economical causes, etc. should be taken into account.

In the aforementioned case, because of partially stated reasons, to impose the carrying capacity of the environment, growth rate $a$ must be taken as[20]:

$$a = a_0 \left(1 - \frac{N_n}{N_{max}}\right) \tag{11}$$

If Eq.(11) is substituted into Eq.(10), for a discrete form of evolution of the population number

$$N_{n+1} - N_n = a_0 \Delta t N_n - \frac{a_0 \Delta t}{N_{max}} N_n^2 \tag{12}$$

is deduced. Eq.(12) can be obtained in a continuous equation form

$$\frac{dN}{dt} = aN - bN^2 \tag{13}$$

by taking $a = a_0$, $b = \frac{a_0}{N_{max}}$ and dividing both sides by $\Delta t$ thereafter taking the limit $\Delta t \to 0$. One can see that Eq.(13) is a logistic equation.

It is observed that, the coefficient $a$ determines the growth of population, while the coefficient $b$ includes the environmental effects and then breaks the population growth. If Eq.(13) is rewritten for $N_{max}$, then

$$\frac{dN_{max}}{dt} = aN_{max} - bN_{max}^2$$

is obtained. In view of the fact that $N_{max}$ is a constant, then $N_{max} = \frac{a}{b}$ is attained.

By an inspection to Eq.(12); for $N_{n+1} > 0$ it requires $\frac{a_0 \Delta t}{1 + a_0 \Delta t} < 1$ and then $N_{max} = \frac{a}{b} \geq N_n$ in all steps.

From discrete cumulative growth equation (12), logistic map can be deduced as

$$N_{n+1} = rN_n - \frac{(r-1)}{N_{max}} N_n^2 \tag{14}$$

where $r = (1 + a_0 \Delta t)$. If the Eq.(14) is rearranged, then for the discrete population number

$$N_{n+1} = rN_n \left(1 - \frac{(r-1)}{rN_{max}} N_n\right) \tag{15}$$

can be written. In order to get a dimensionless population expression

$$N_n = \frac{rN_{max}}{r-1} x_n \tag{16}$$

is choosen. Hence, Eq.(16) is substituted into Eq.(15), then the logitic map

$$x_{n+1} = rx_n(1 - x_n) \tag{17}$$

is obtained[3,4]. Thus, the logistic map which is the simplest nonliner discrete equation which arise in many context is related with the cumulative processes. As a result, logistic map is a nonlinear, dimensionless equation that accommodate the limitation of cumulative growth naturally.

**4.Cumulative approach to population dynamics of a community**

Especially, in devoloping countries, the migration of the rural population to urban is frequently observed. Consequently, ghettos and the gecekondus/shanty houses are surround the cites. In the general sense, the population growth rate of this migrated component of population, which live around cities in isolation, in fact, is rather bigger than the native, urban population growth rate. Also, due to several different reasons, refugee movements are observed between countries as well. In our time, due to increasing mobility of human communities(refugees), it is rather important to compare the amounts of currently living population and the population which is incorporated to them by the colonies in a geography. For this purpose, to follow the growth of two heterogeneos components of population, such a scenario is proposed. Let us suppose that, in every time interval $\Delta t$, a colonial society $P$ joins to native(primary) population $N_0$ that lives in a zone. Here it is supposed that both components of the population namely, native and colonial respectively have equal growth rates within the framework of cumulative approach. Let us investigate, the accumulative population $K$. Here, there are two cases that can occur for the accumulative population: summation of the native and colonial population or subtraction of them.

First of all, to compare the two components of the population, we will consider the difference of them.

Let us handle the case where population attains the value after $n$ steps, during time $t$. Thus, for every discrete time interval $\Delta t$, corresponding successive steps are as follows:

$0^{th}$ step $\qquad K_0 = \phi^0 N_0$

$1^{st}$ step $\qquad N_1 = N_0 + a\Delta t N_0 - P = (1 + a\Delta t)N_0 - P = \phi N_0 - P$

$2^{nd}$ step $\qquad N_2 = \phi N_1 - P = \phi^2 N_0 - (\phi + 1)P$

$\qquad\qquad\qquad \vdots \qquad \vdots \qquad \vdots$

$n^{th}$ step $\qquad N_n = \phi N_{n-1} - P = \phi^n N_0 - (\phi^{n-1} + \cdots + \phi + 1)P \qquad (18)$

According to this equation, accumulative population can be written symbolically as

$$K = L - T \qquad (19)$$

where,

$$L = \phi^n N_0 \qquad (20a)$$

and

$$T = P \sum_{k=0}^{n-1} \phi^k = P \frac{1-\phi^n}{1-\phi}. \tag{20b}$$

Thus, for Eq. (19)

$$K = \phi^n N_0 - P \frac{1-\phi^n}{1-\phi} \tag{21}$$

is obtained[14,15].

For the accumulative population K, the case $L > T$ corresponds to the positive $K$, where primary population growth exceeds the immigrant population growth. The case $L = T$, namely $K = 0$, means that the primary population growth is in equilibrium with immigrant population growth. In the case $L < T$ namely $K < 0$, it is understood that, primary population growth is less than the immigrant population. It should be emphasised that comparison of the immigrant population and native urban population depends on two variables namely growth rate $a$ and time $t$.

Let us illustrate our model by chosen quantities. Consider that the accumulative population $K$ attain the value

$$K_\lambda = \lambda N_0 \tag{22}$$

where $\lambda$ is a rational number. When Eq.(22) is substituted in Eq. (21), then $P_\lambda$ is calculated as

$$P_\lambda = N_0 \frac{(\lambda - \phi^n)(\phi - 1)}{1 - \phi^n}. \tag{23}$$

As an application, for the values $\lambda = 1, 0, -1$, corresponding $P_\lambda$ quantities are

$$P_1 = N_0(\phi - 1) = \frac{N_0}{n} at \tag{24}$$

$$P_0 = N_0 \frac{\phi^n(1-\phi)}{1-\phi^n} \tag{25}$$

and

$$P_{-1} = N_0 \frac{(1+\phi^n)(1-\phi)}{1-\phi^n} \tag{26}$$

respectively.

In the case $\lambda = 1$, accumulated population remains unchanged at the end of time $t$. In the case $\lambda = 0$, accumulated population tends to zero in the course of time. In the case $\lambda = -1$, the reached immigrant population exceeds native population by an amount $N_0$.

By the inspiration of nonextensive thermostatistical dynamics[23], let us determine the total population that is composed of the populations $L$ and $T$ together. Until now, $L$ and $T$ populations are treated as independent quantities of the system. At this stage, we confront to nonadditivity of populations .

In fact, the components of populations, naturally interact socially, economically,etc. that live in the same geography. When, populations $L$ and $T$ are interdependent, after some time $t$,accumulative population $K_q$ can be written as

$$K_q = L_q + T_q + (1-q)L_q T_q. \tag{27}$$

It is due to this property that for $q \neq 1$, $K_q$ is said to be nonadditive, while, for $q = 1$, $K_1$ is additive where $L_1$ and $T_1$ do not interact.

Since $K_q$ is a nonnegative quantity, it follows that, for independent subsystems $L_q$ and $T_q$, $K_q \geq L_q + T_q$, if $q < 1$ and $K_q \leq L_q + T_q$ if $q > 1$. Consistently, the cases $q < 1; K_{q<1}$ and $q > 1; K_{q>1}$ may be referred as the superadditive and subadditive respectively. It may be concluded that $q$ is a measure of social interaction between the societies which could be named as social interaction index. Up to Eq. (27), when $q = 1$, there is no interaction between the two components, in this case the accumulative population is additive, then from Eq. (20a) and Eq.(20b)

$$K = \phi^n N_0 + P \frac{1-\phi^n}{1-\phi} \tag{28}$$

it can be written. As a special scenario, when the immigrant population is equal to primary population, that is $P = N_0$, the relation

$$K = N_0 \frac{1-\phi^{n+1}}{1-\phi} \tag{29}$$

is deduced.

## 5.Differential equation of density function for immigrant population

It may be interesting to consider the balance of primary population and the colonial population after a time $t$. Let us examine the realization of equilibrium situation mathematically in terms of physical quantities. In the equilibrium state, where the primary and colonial populations are equal, the following equation

$$P(z) = \frac{z}{1-\left(1+\frac{z}{n}\right)^{-n}} \frac{N_0}{n} \tag{30}$$

can be written from Eq.(25),where $z = at$. Hence,$P$ can be written in terms of $\frac{N_0}{n} = n_0$ as a probability function $f(z)$. Thus, for stable case, probability density $f(z)$

$$f(z) = \frac{z}{1 - \left(1 + \frac{z}{n}\right)^{-n}} \tag{31}$$

can be written. For sufficiently large values of $n$, function $f(z)$ behaves as

$$f(z) = \frac{z}{1 - e^{-z}}. \tag{32}$$

Let us derive the differential equation for the probability function $f(z)$ in the state of equilibrium. For this purpose, with the help of the generating function of Bernoulli numbers,

$$f(z) = \frac{z}{1 - e^{-z}} = \sum_{m=0}^{\infty} B_m \frac{(-z)^m}{m!} \tag{33}$$

can be written, where $B_m$ are the Bernoulli numbers[21]. Using the values of the first few Bernoulli numbers $B_0 = 1$, $B_1 = -\frac{1}{2}$, $B_2 = \frac{1}{6}$, $B_4 = -\frac{1}{30}$, $B_6 = \frac{1}{42}$, ..., probability density can be written as

$$f(z) \approx 1 + \frac{1}{2}z + \frac{1}{6}z^2 - \frac{1}{30}z^4 + \frac{1}{42}z^6 + \cdots \tag{34}$$

Thus, a differential equation can be derived for the state of equlibrium, whose solution is $f(z)$. Then, from the relations between the quantites $f(z)$, $f^2(z)$ ve $f'(z)$, the following differential equation can be established in terms of density function $f(z)$

$$f'(z) + p(z)f(z) = q(z)f^2(z) \tag{35}$$

where,

$$p(z) = -\frac{1}{z}, \qquad q(z) = -\frac{e^{-z}}{z}. \tag{36}$$

As it can be recognized that, Eq.(35) is formally a Bernoulli equation[3,4]. This means that the case of equilibrium where the components of population, respectively native and colonial are equal, there exist a first order nonlinear equation. The solutions of Eq.(35) are naturally the probability function $f(z)$, which is given in Eq. (32) and Eq.(34). Here, it is interesting to observe that the solution can be written as a series expansion in terms of Bernoulli numbers. With this regard, Bernoulli numbers are confronted as the generator of the solution of a differential equation describing population growth of a community.

## 6.Investigation of population growth in the framework of fractional calculus

The essential mechanism underlying the success of fractional calculus in solving differential equation is discussed in our previous works[14,15]. In this regard, since population growth evolve in a non-Euclidean, fractal and fractured space, and in a non-Markovian manner, this process must be handled in the framework of fractional calculus. In this respect, fractional rate equation can be written for the population gowth as

$$D_t^\alpha N(t) - N_0 \frac{t^{-\alpha}}{\Gamma(1-\alpha)} = a^\alpha N(t). \qquad (37)$$

If the solution of Eq.(37) is handled with the Laplace transformation method, then

$$N(t) = N_0 E_\alpha (at^\alpha) \qquad (38)$$

is attained[9,14,15,18].

It is observed that, Eq.(4), which is obtained in section1 by cumulative growth approach for the population growth is same as the result which is obtained in the framework of fractional calculus, given in Eq.(38). The reality is put forwarded that the essential mechanism underlying in the success of fractional calculus is cumulative growth. In the case $\alpha \to 1$, Eq.(38) is reduced to standard Malthus equation; Eq.(8).

## 7. Investigation of population dynamicsof a community with two components using fractional calculus

In this section, let us reconsider the population dynamics in the framework of fractional calculus. Let the evolution operator be given as $\phi = (1 + a\Delta t)^\alpha$. Here, the parameter $\alpha$ will be estimated as a measure of environmental effect to the population evolution. In fractional calculus, parameter $\alpha$ is identified as the order of differintegration. Let us now, consider the discrete process, which evolves in every discrete successive time interval $\Delta t$ cumulatively. Then, accumulative population $K$ of the above scheme is formed as:

$0^{th}$ step  $\quad K_0 = (\phi)^0 N_0$

$1^{st}$ step  $\quad K_1 = \phi N_0 - P$

$2^{nd}$ step  $\quad K_2 = \phi^\alpha N_1 - P = (\phi)^2 N_0 - [\phi + 1]P$

$\quad \vdots$

$n^{th}$ step  $\quad K_n = (\phi)^n N_0 - [(\phi)^{n-1} + (\phi)^{n-2} + \cdots + \phi + 1]P \qquad (39)$

From Eq.(39),

$$K = (\phi)^n N_0 - P \sum_{k=0}^{n-1} (\phi)^k \qquad (40)$$

can be written. In the equilibrium state these primary and colonial populations are equal thus

$$K = L - T = 0 \qquad (41)$$

can be written from Eq.(40), for accumulated population. In Eq.(40), if the summation on the second term is performed, then for the immigrant population $P$,

$$(\phi)^n N_0 = P \frac{1-(\phi)^n}{1-\phi^\alpha} \tag{42}$$

can be written and from Eq.(42)

$$P = \frac{\phi - 1}{1 - \phi^{-n}} N_0 \tag{43}$$

is deduced. For $\Delta t = t/n$,

$$P = \frac{(1 + a\frac{t}{n})^\alpha - 1}{1 - (1 + a\frac{t}{n})^{-\alpha n}} N_0 \tag{44}$$

can be written. From the Eq.(44), for $\frac{n}{N_0} = n_0$, the quantity $P$, can be expressed in terms of $n_0$ as a probability function $\rho(t)$. Thus, for $a\frac{t}{n} \ll 1$, the probability function is

$$\rho(t) = \frac{\alpha a t^\alpha}{1 - E_\alpha(at^\alpha)} . \tag{45}$$

A differential equation whose solution is given by $\rho(t)$ can be written for the equilibrium state. For this purpose, a fuction $v(at^\alpha)$ can be defined as

$$\frac{\alpha a t^\alpha}{\rho(t)} = 1 - E_\alpha(at^\alpha) = v(at^\alpha) . \tag{46}$$

To proceed further, let us use the definition of derivative of M-L function which is given by

$$D_{0+}^\alpha[E_\alpha(at^\alpha)] = \frac{t^{-\alpha}}{\Gamma(1-\alpha)} + aE_\alpha(at^\alpha) \tag{47}$$

where $D_{0+}^\alpha$ is Riemann-Liouville derivative[16,17]. Then, from Eq.(47), a fractional logistic equation can be deduced for the function $v(az^\alpha)$ given by Eq.(46) in the following form;

$$D_{0+}^\alpha[v(at^\alpha)] = a[v(at^\alpha) - 1] . \tag{48}$$

Here, a compoundly growing population which is composed of two components and which takes into account environmental effects is considered in the state of equilibrium. Then, for this equilibrium state, a fractional differential equation which is given by Eq.(48) is obtained. In other words, solutions of the equation which is given by Eq.(48), exhibit the evolution of the immigrant population $P$.

## 8. Hybrid solution of the logistic equation

In this section, let us consider the cumulative evolution of a population growth for which growth rate is restricted. In order to find the solution of logistic equation of population given in Eq.(13) we reconsider the problem within the framework of cumulative growth process. In Eq.(13), there are(emerge) four cases with respect to algebraic sign of the coefficients $a$ and $b$ which are; $a > 0$ and $b > 0$, $a > 0$ and $b < 0$, $a < 0$ and $b > 0$, $a < 0$ and $b < 0$ individually that determines the character of the biodynamics. The cases are examined in the following discussion. For the equilibrium state, the coefficients $a$ and $b$ in Eq.(13) are choosen as positive and constant. The sign of these coefficients in Eq.(13) are determined within the course of the process. Namely, if the sign of $a$ is taken positive, it corresponds to the growth of population at the beginning of the process. If the sign of $a$ is taken negative, it corresponds to diminution of a population. Consideration of the sign of $b$ as positive, gives rise to the growth process to slow downas time progresses that is breaking effect, otherwise supports the growing of population.

To solve Eq.(13), let us havethe following transformation

$$N(t) = \frac{1}{u(t)} \tag{49}$$

then the Eq.(13)is reducedto

$$\frac{du(t)}{dt} + au(t) = b. \tag{50}$$

Eq.(50) is a first order inhomogeneous lineer differential equation[3,4] which is known as rate equation that can be solved by two methods namely fractional calculus approach and cumulative diminution method respectively. To observe the mechanism behind the success of fractional mathematics and to compare the solutions, let us consider the solution of Eq.(50) by means of cumulative diminution method. The solution of the homogeneous part of Eq.(50) is given [14,15] in terms of M-L function as

$$u_h = cE_\alpha(-at^\alpha) \tag{51}$$

which is obtained byusing cumulative diminutions in which diminution operator $\phi = (1 - a\Delta t)^\alpha$ is involved.

The general solution of Eq.(50) consists of the summation of the homogeneous solution and the particular solution. If the particular solution is taken as $u_p = b/a$, then the general solution is deduced

$$u_g = cE_\alpha(-at^\alpha) + \frac{b}{a} \tag{52}$$

where $a, b, c$ are the constants which are determined by the initial condition and limiting values. In view ofthe transformation (49), the population $N(t)$ is obtained as

$$N(t) = \frac{1}{cE_\alpha(-at^\alpha) + \frac{b}{a}}. \tag{53}$$

By taking into account the initial population $N(0) = N_0$ at $t = 0$, Eq.(53) can be explicitly written as

$$N(t) = \frac{N_0}{\left(1 - \frac{b}{a}N_0\right)E_\alpha(-at^\alpha) + \frac{b}{a}N_0}. \tag{54}$$

It should be underlined that, population $N(t)$ strictly depends on initial conditions which are determined with respect to equilibrium state $\frac{a}{b}$. Morever, the development of the population growth or diminution depends on the signs of fertility rate $a$ and breaking term $b$. The term $-bN^2(t)$ in Eq.(13) prevents the population from growing boundless. The solution in Eq.(54) is plotted in Fig.2.

If the initial condition is in interval $0 < N_0 < \frac{a}{b}$, then population function increases monotonically and in the long time limit it reaches to its equilibrium value $N_{max} = \frac{a}{b}$ which is stated in section2.

On the other hand, if $N_0 > \frac{a}{b}$, then the population $N(t)$ decreases monotonically and in the long time limit reaches to the value $N_{max} = \frac{a}{b}$.

As a special case, for $b = 0$, the solution (53) reduces to following form

$$N(t) = \frac{1}{c\, E_\alpha(-at^\alpha)} \tag{55}$$

which can be named as modified Malthus law[1].

For the benefit of scientist who are interested in the prediction of population developments by the application of regression analysis Eq.(53) is rewritten as

$$N(t) = \frac{A}{B\, E_\alpha(-at^\alpha) + 1} \tag{56}$$

where A stands for $\frac{a}{b}$ and B stands for $\frac{ca}{b}$.

If the environmental effects are ignored, namely for $\alpha \to 1$, the growth of a population is expressed as[3,4]

$$N(t) = \frac{N_0}{\left(1 - \frac{b}{a}N_0\right)\exp(-at) + \frac{b}{a}N_0}. \tag{57}$$

If the Eq. (54) is determined by using the limit properties of M-L function which are given in Eq.(6a), then for the $N(t)$

$$N(t) \approx N_0 \tag{58}$$

can be written when $at \ll 1$. Then, from the limit of M-L function Eq.(6b), namely for $at \gg 1$,

$$N(t) \approx \frac{a}{b} \tag{59}$$

is obtained[9,12,19].

It is observed that the carrying capacity is independent of time and initial values. As a conclusion, quantity $\frac{a}{b}$ is the maximum population namely maximum carrying capacity that system can attain at the equilibrium state in the long time limit.

It is concluded that the initial value is either larger than the carrying capacity value, namely $N_0 > \frac{a}{b}$ for the diminishing system, or less than the carrying capacity value namely $0 < N_0 < \frac{a}{b}$ for the growing sytem, eventually the system leads to the equilibrium value.

When the environment effect of the system is ignored, $b = 0$ is taken then modified Malthus law Eq.(55) is reduced to historical well known Malthus law

$$N(t) = N_0 \exp(at).$$

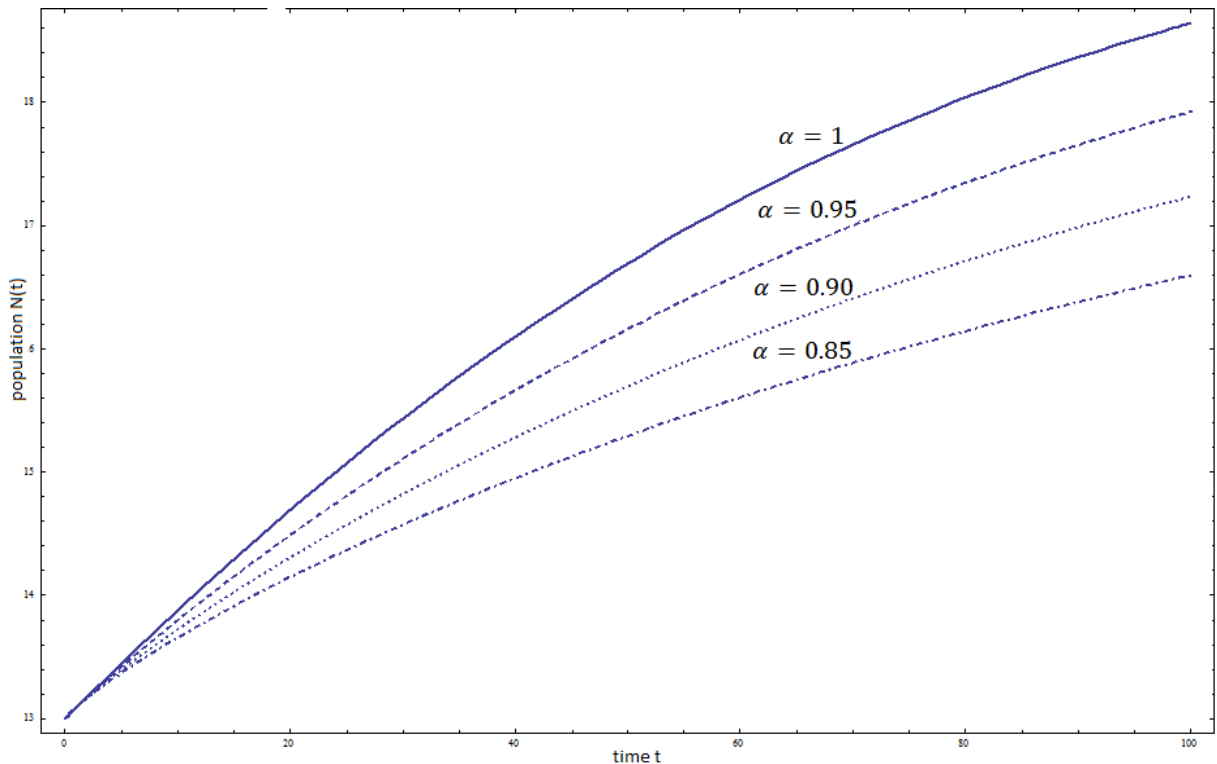

Fig.2. The graph of hybrid solution of logistic equation; Eq.(54) for $0 < N_o < \frac{a}{b}$ for different values of $\alpha$. It is plotted for the initial population $N_0 = 13$ units, fertility rate $a = 0.02$, $b = 0.001$ and $\alpha = 1$ (bold), $\alpha = 0.95$ (dashed), $\alpha = 0.9$ (dotted), $\alpha = 0.85$ (dotdashed)

In Fig.2., the curves for solutions of the expression which is given in Eq.(54) are plotted for the different values of $\alpha$. According to Fig.2. it can be observed that as the values of $\alpha$ increase, it fastens the population growth and then population attains the equilibrium. However, if the hybrid solution which is obtained by cumulative approach is compared with the standard solution of logistic equation that is the case $\alpha = 1$, it is observed that the parameter $\alpha$ acts as a measure of the environmental effect which slows down the growth process and behaves like a retardation effect.

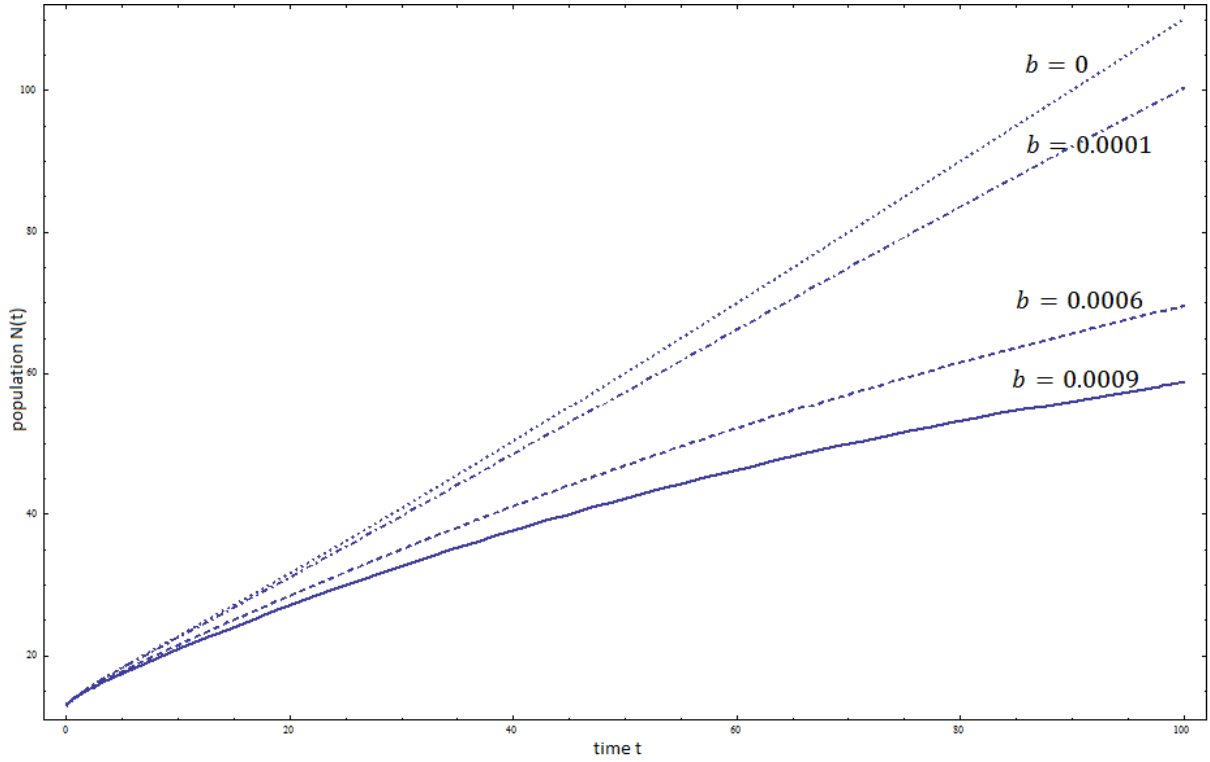

Fig.3.The graph of hybrid solution of logistic equation for $0 < N_o < \frac{a}{b}$, with different values of $b$. It is plotted for the initial population $N_0 = 13$ units, fertility rate $a = 0.1$, $\alpha = 0.75$ and $b = 0$ (dotted), $b = 0.0001$(dotdashed), $b = 0.0006$ (dashed), $b = 0.0009$ (bold).

In Fig.3, solution of $N(t)$ given by Eq.(54)is plotted for different values of $b$ . For the increasing values of $b$, it can be observed that the slowdown and breaking effects are imposed on exponential growth .

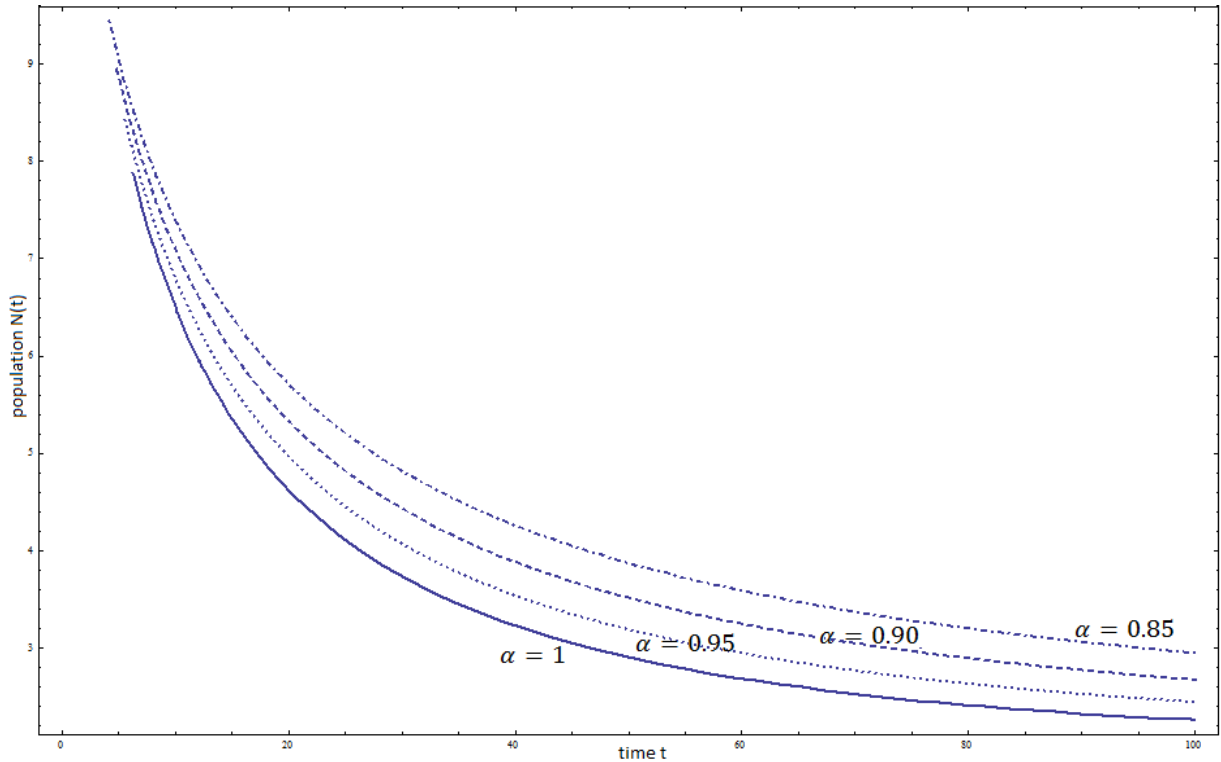

Fig.4. The graph of hybrid solution of logistic equation for $N_0 > \frac{a}{b}$, for different values of $\alpha$. It is plotted for the initial population $N_0 = 13$ units, fertility rate $a = 0.02$, $b = 0.01$ and $\alpha = 1$ (bold), $\alpha = 0.95$ (dotted), $\alpha = 0.9$ (dashed), $\alpha = 0.85$ (dotdashed)

In Fig.4., the curves of Eq.(54) for the case $N_0 > \frac{a}{b}$ are plotted for the different values of $\alpha$. According to Fig.4. it can be observed that as the values of $\alpha$ increase, it fastens the population diminution.

In the above solution, sign of $a$ and $b$ are choosen as $a > 0$ and $b > 0$ then the stability of the system is discussed. Algebraically, in Eq.(13), there could be two choices, namely, sign of $a$ positive, $b$ negative or signs of $a$ and $b$ are both negative. For the former case, besides the contribution of the $a$, population proceed to grow with the contribution of the environmental effects due to $b$. On the other hand, in the latter case, the first term slows down the population while the second term causes growing effect. But, due to the reason that in both cases the environmental effects which are taken into account by thesecond term $b$, run parallel to the first term, consequantly the breaking term $b$ looses its effect on equilibrium. Thus, the carrying capacity has not been taken into account. As a result, the choice of signs of parameters $a$ and $b$ and concomitant equations are not acceptable for the population dynamics.

In addition to above choices, lastly, negatif $a$ and positive $b$ could be choosen, thus both of the terms causes a diminishing effect on the population and then makes the system to develop faster to extinction of the species. According to last choice, from Eq. (53), the population evolution can be written by

$$N(t) = \frac{N_0}{(1 + \frac{b}{a}N_0)E_\alpha(at^\alpha) - \frac{b}{a}N_0}. \tag{60}$$

Diminishing population function $N(t)$ is reduced[3,4] into following form:

$$N(t) = \frac{N_0}{(1 + \frac{b}{a}N_0)exp\,(at) - \frac{b}{a}N_0} \tag{61}$$

in the case $\alpha \to 1$[3,4].

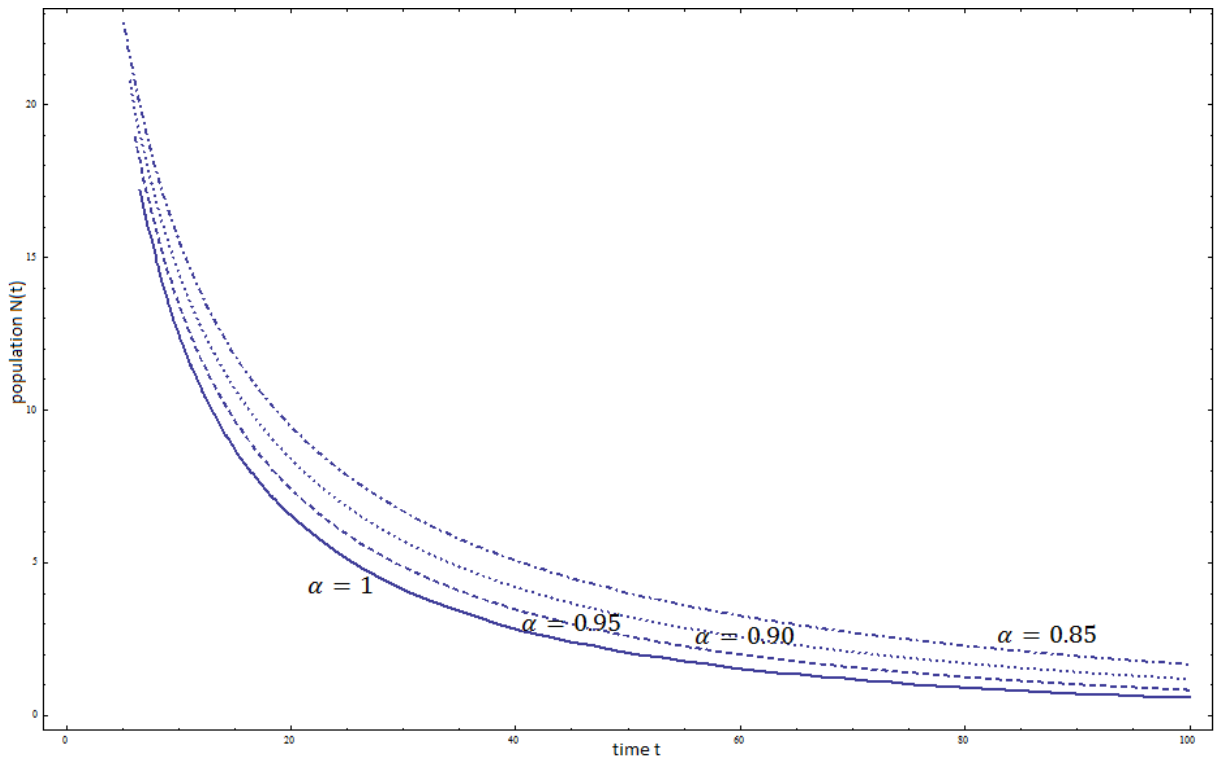

Fig.5.The figure of hybrid solution of logistic equation for $a < 0$ and $b > 0$, for different values of $\alpha$. It is plotted for the initial population $N_0 = 50$ units, fertility rate $a = -0.02$, $b = 0{,}005$ and $\alpha = 1$ (bold) , $\alpha = 0.95$(dashed), $\alpha = 0.9$(dotted) , $\alpha = 0.85$(dotdashed).

In Fig.5., Eq.(60) are plotted for the different values of $\alpha$. It is observed that as the values of $\alpha$ increase, it fastens the population diminution which lead to extinction of the examined species.

In the solution (60), the case $b = 0$ corresponds to the situation, where the breaking effect is absent and exponential diminution is prevailed.

**9.The results and discussions**

In this study, population dynamics is reviewed since the human population movements are increasing nowadays in the world. Thus, it becomes important for the governments to predict the attained value of populations in future, to be able to plan and to cope with the social problems of future. From this perspective, cumulative growth which is inspired by the spirit of Fibonacci approach and the fractional calculus formulations are used in parallel in order to solve the equations of population dynamics. In this content, historical formulas, concerning to population dynamics, especially Malthus and Verhulst equations are reconsidered and updated. In the fractional calculus formulation, interaction of population with its environment is imposed to the dynamics with the order of differentiation $\alpha$, which is defined in the interval [0,1]. On the other hand, by discrete cumulative diminishing/growth approach, by taking into account the fertility rate limitations due to the carrying capacity of environment, ultimately logistic equation is deduced. The logistic equation is derived through our cumulative growth approach thus the cycling mechanism of biodynamics behind the logistic equation is uncovered.

A scenario concerning to the nowadays population movements is constructed and examined in detail. Due to its importance, the attained quantities of growths of native population and immigrant population in a regionare formulated and the results are compared. The development of the components of the population are illustrated. The equilibrium of the two components of population are examined in detail. In this content, for the state of equilibrium between the native and the immigrant populations, the value of colony population $P$ is obtained in terms of growth rate $a$, initial population $N_0$ of native population and the order of differentiation $\alpha$ which involves environmental effect. An immigrant population density function $f(z)$ is obtained and described in terms of the Bernoulli numbers. Time inclusive differential equations which exhibit the evolution of the colonial population are obtained in the framework of cumulative growth and fractional calculus approaches, where the density function $f(z)$ is a solution.

The differential equation for a population $N(t)$ which is obtained within the framework of cumulative growth which takes into account the environmental effects is solved hybridly. In the long time limit, the hybrid solution of population which interacts with ecosystem,evolve a limiting value that is independent of the initial value of population.It is concluded that, the initial value of population is less or more than the carrying capacity,in the long time limit, the population tends to carrying capacity of the ecosystem.

When the system arrives to the equilibrium state, the population fluctuate around the carrying capacity. The relation between the hybrid solution which is put forward in this study and other well known solution of historical equations are investigated by means of plotted graphs. In the cumulative approach where environmental effects have been taken into account, the function that describes the population dynamics, M-L function appears as an ubiquitous function. As a conclusion, in population dynamics, the population growth parameter $a$ is the main parameter that determinesthe growth of the population whereas the parameter $b$ takes care of carrying capacity by performing breaking effect, eventually the system leads to the equilibrium state.

In this work, the population of a community which is composed of two components is determined. It is obvious that each of these components are liable to carrying capacity limitations. These population dynamicsof the community are handled in the framework of the cumulative growth and the fractional calculus approaches. For the simplicity of the mathematical formulations, growth rates of the components of the community that are named as native which live in urban and immigrant originating from rural are considered as equal. In reality, it is known that the growth rate of the immigrant population is larger than the native population growth rate.By taking into account the interaction of the components of community, nonadditive property of population of the components has been introduced. In this context, the social interaction index $q$ which is a measure of interaction of the two components of the community is inserted.

In this study, in order to infer information about the mechanism of biodynamics, physical approaches of cumulative growth and the mathematical approaches of fractional calculus are used in parallel to describe the discrete and continuous population dynamics. It is concluded that the solution of the differential equation of population dynamics incorporates two different opposite effects, one of them is the first term which increases population number while the second term decreases the population number in the course of time and eventually leads to an equilibrium state which takes care of the carrying capacity. Thus two opposite effects are reconciled in order to maintain life in harmony.